\documentclass[3p,times,procedia]{elsarticle}

\usepackage{ecrc}

\volume{00}

\firstpage{1}

\journalname{Physics Procedia}

\runauth{}

\jid{phpro}

\jnltitlelogo{Physics Procedia}

\CopyrightLine{2011}{Published by Elsevier Ltd.}

\usepackage{amssymb}
\usepackage[figuresright]{rotating}

\usepackage{appendix}
\usepackage{array}
\newcolumntype{x}[1]{>{\centering\arraybackslash}p{#1}}

\usepackage{epsfig}
\usepackage{graphicx}
\usepackage{dcolumn}
\usepackage{amsmath}
\usepackage{enumerate}
\usepackage{mathrsfs}

\usepackage{color}
\usepackage{ulem}

\newcommand{\eg}{e.g.~}

\newcommand{\beq}{\begin{equation}}
\newcommand{\eeq}{\end{equation}}
\newcommand{\ud}{\text{d}}

\newcommand{\mDM}{m}
\newcommand{\ER}{E_\text{R}}
\newcommand{\Ed}{E'}
\newcommand{\vesc}{v_\text{esc}}
\newcommand{\vmin}{v_\text{min}}

\newcommand{\eR}{\mathcal{R}}

\newcommand{\bol}[1]{\boldsymbol{#1}}
\newcommand{\bfv}{\bol{v}}
\usepackage{color}
\definecolor{rossoCP3}{cmyk}{0,.88,.77,.40}
\definecolor{verdeCP3}{rgb}{0.09765625, 0.57421875, 0.1015625}
\definecolor{bluCP3}{rgb}{0, 0.23, 0.67}

\ifx\pdfoutput\undefined
\usepackage[dvips,bookmarks]{hyperref}    % This is for arXiv.org
\else
\usepackage{hyperref}    % This is for pdftex
\fi
\hypersetup{colorlinks, bookmarksopen, bookmarksnumbered,
citecolor=verdeCP3, linkcolor=bluCP3, pdfstartview=FitH, urlcolor=rossoCP3}

\begin{document}

\begin{frontmatter}



\dochead{}
%% e.g. \dochead{17th International Conference on Dynamical Processes in Excited States of Solids}

\title{Update on the Halo-Independent Comparison\\of Direct Dark Matter Detection Data}


\author[a]{Eugenio Del Nobile\corref{nota}}
\ead{delnobile$@$physics.ucla.edu}
\author[a]{Graciela B.~Gelmini}
%\ead{gelmini$@$physics.ucla.edu}
\author[b]{Paolo Gondolo}
%\ead{paolo.gondolo$@$utah.edu}
\author[a]{Ji-Haeng Huh}
%\ead{jhhuh$@$physics.ucla.edu}
\address[a]{Department of Physics and Astronomy, UCLA, 475 Portola Plaza, Los Angeles, CA 90095, USA}
\address[b]{Department of Physics and Astronomy, University of Utah, 115 South 1400 East \#201, Salt Lake City, UT 84112, USA}

\cortext[nota]{Speaker}

\begin{abstract}
We briefly review the halo-independent formalism, that allows to compare data from different direct dark matter detection experiments without making assumptions on the properties of the dark matter halo. We apply this method to spin-independent WIMP-nuclei interactions, for both isospin-conserving and isospin-violating couplings, updating the existing analyses with the addition of the SuperCDMS bound. We point out that this method can be applied to any type of WIMP interaction.
\end{abstract}


\end{frontmatter}

% \linenumbers


\section{Introduction}

At present, four direct dark matter (DM) search experiments (DAMA \cite{Bernabei:2010mq}, CoGeNT \cite{Aalseth:2010vx, Aalseth:2011wp, Aalseth:2012if, Aalseth:2014eft, Aalseth:2014jpa}, CRESST-II \cite{Angloher:2011uu}, and CDMS-II-Si \cite{Agnese:2013rvf}) have data that may be interpreted as signals from DM particles in the light WIMPs (for weakly interacting massive particles) range. DAMA \cite{Bernabei:2010mq} and CoGeNT \cite{Aalseth:2011wp, Aalseth:2014eft, Aalseth:2014jpa} report annual modulations in their event rates, compatible with those expected for a DM signal \cite{Drukier:1986tm, Freese:2012xd}. CoGeNT \cite{Aalseth:2010vx, Aalseth:2012if, Aalseth:2014eft, Aalseth:2014jpa}, CRESST-II \cite{Angloher:2011uu}, and CDMS-II-Si \cite{Agnese:2013rvf}, observe an excess of events above their expected backgrounds, that may be interpreted as due to DM WIMPs.

However, other experiments do not observe significant excesses above their estimated background, thus setting upper limits on the interaction of WIMPs with nuclei. The most stringent limits on the average (unmodulated) rate for light WIMPs are set by the LUX \cite{Akerib:2013tjd}, XENON10 \cite{Angle:2011th}, XENON100 \cite{Aprile:2012nq}, CDMS-II-Ge \cite{Ahmed:2010wy}, CDMSlite \cite{Agnese:2013jaa} and SuperCDMS \cite{Agnese:2014aze} experiments, with the addition of SIMPLE~\cite{Felizardo:2011uw}, PICASSO \cite{Archambault:2012pm} and COUPP \cite{Behnke:2012ys} for spin-dependent and isospin-violating interactions. CDMS-II-Ge \cite{Ahmed:2012vq} also constrains directly the amplitude of an annually modulated signal.

In order to compare a model for WIMPs with data from direct DM detection experiments, one needs to assume a value for the DM local density and velocity distribution in our galaxy. The Standard Halo Model (SHM) is usually assumed for the DM halo, corresponding to a truncated Maxwell-Boltzmann distribution for the DM velocity (see \eg \cite{Savage:2008er}). However, the parameters of this model are not known to great accuracy, and the model itself is not supported by data. Actually, quantitatively different velocity distributions are obtained from numerical simulations (see \eg \cite{Mao:2012hf}). Various models and parametrizations for the DM velocity distribution in our galaxy have been proposed as alternatives to the SHM, either derived from astrophysical data or from N-body simulations (see \eg \cite{Freese:2012xd} and references therein). Other authors have attempted to estimate the uncertainty in the determination of the properties of the DM halo, and to quantify its effects on the interpretation of DM direct detection data (see \eg \cite{Green:2002ht, McCabe:2010zh, Green:2010gw, Green:2011bv, Fairbairn:2012zs}). Another approach is that of marginalizing over the parameters of the DM halo when computing bounds and allowed regions from the experimental data (see \eg \cite{Arina:2013jma}). However, all these procedures maintain a certain degree of model dependence, \eg in the choice of the functional form of the parametrization of the halo. It is very important to notice here that the high velocity tail of the DM velocity distribution plays a crucial role in determining the number of DM particles that are above threshold for a given experiment, and therefore a way to analyze the data without the need to make any assumption on its shape is highly desirable.

The problem of comparing results from different direct detection experiments can indeed be formulated without the need to assume a velocity profile for the DM \cite{Fox:2010bz, Frandsen:2011gi, Gondolo:2012rs, Frandsen:2013cna, DelNobile:2013cta, DelNobile:2013cva, DelNobile:2013gba, DelNobile:2014eta, Feldstein:2014gza, Fox:2014kua} (see also \cite{HerreroGarcia:2011aa, HerreroGarcia:2012fu, Bozorgnia:2013hsa}). The basic idea is to factor out from the formulas used to compute the scattering rate, all the astrophysical quantities such as the DM velocity distribution function. In this way the rate can be computed, for any model of particle interactions between the DM and the nuclei in the detector, with no need to assume a velocity profile for the DM, while rather allowing to use the experimental data to constrain the unknown quantities. Such a ``halo-independent'' analysis is particularly useful to investigate the compatibility of the different experimental results in the light WIMP hypothesis, for which the details of the DM velocity distribution, especially at high velocities, are notably relevant.

Here we summarize the results presented in \cite{DelNobile:2013cta, DelNobile:2013gba} for spin-independent (SI) interactions with both isospin-conserving and isospin-violating \cite{Kurylov:2003ra, Feng:2011vu} couplings, and update them with the addition of the SuperCDMS \cite{Agnese:2014aze} results. For the data analysis we follow \cite{DelNobile:2014eta}, except for the CoGeNT 2014 halo-independent analysis for which we follow \cite{DelNobile:2013gba}. The halo-independent method can be applied to any kind of WIMP-nucleus interaction \cite{DelNobile:2013cva}, \eg WIMPs with a magnetic dipole or an anapole moment \cite{DelNobile:2013cva, DelNobile:2014eta}.

\section{Scattering rate for spin-independent interaction}

What is observed at direct DM detection experiments is the WIMP-nucleus differential scattering rate, usually measured in units of counts/kg/day/keV. For a target nuclide $T$ initially at rest, recoiling with energy $\ER$ after the scattering with a WIMP with mass $m$ and initial velocity $\bfv$, the differential rate is
\beq
\label{dRdER}
\frac{\ud R_T}{\ud E_\text{R}} = \frac{\rho}{\mDM} \frac{C_T}{m_T} \int_{v \geqslant v_\text{min}(\ER)} \hspace{-24pt} \ud^3 v \, f(\bfv, t) \, v \, \frac{\ud \sigma_T}{\ud \ER}(\ER, \bfv) \ .
\eeq
Here $m_T$ is the target nuclide mass and $C_T$ is its mass fraction in the detector, and we denoted with $v = | \bfv |$ the WIMP speed. $\ud \sigma_T / \ud \ER$ is the differential scattering cross section. The dependence of the rate on the local characteristics of the DM halo is contained in the local DM density $\rho$ and the DM velocity distribution in Earth's rest frame $f(\bfv, t)$, which is modulated in time due to Earth's rotation around the Sun \cite{Drukier:1986tm, Freese:2012xd}. The distribution $f(\bfv, t)$ is normalized to $\int \ud^3 v \, f(\bfv, t) = 1$. In the velocity integral, $\vmin(\ER)$ is the minimum speed required for the incoming DM particle to cause a nuclear recoil with energy $\ER$. For an elastic collision
\beq\label{vmin}
\vmin = \sqrt{\frac{m_T \ER}{2 \mu_T^2}} \ ,
\eeq
where $\mu_T = m \, m_T / (m + m_T)$ is the WIMP-nucleus reduced mass.

To properly reproduce the recoil rate measured by experiments, we need to take into account the characteristics of the detector. Most experiments do not measure the recoil energy directly but  rather a detected energy $\Ed$, often quoted in keVee (keV electron-equivalent) or in photoelectrons. The uncertainties and fluctuations in the detected energy corresponding to a particular recoil energy are expressed  in a (target nuclide and detector dependent) resolution function $G_T(\ER, \Ed)$, that gives the probability that a recoil energy $\ER$ (usually quoted in keVnr for nuclear recoils) is measured as $\Ed$. The resolution function incorporates the mean value $\langle \Ed \rangle = Q_T \ER$, which depends on  the energy dependent quenching factor $Q_T(\ER)$, and the energy resolution $\sigma_{\ER}(\Ed)$.  Moreover, experiments have one or more counting efficiencies or cut acceptances, denoted here as $\epsilon_1(\Ed)$ and $\epsilon_2(\ER)$, which also affect the measured rate. Thus the nuclear recoil rate in eq.~\eqref{dRdER} must be convolved with the function $\epsilon_1(\Ed) \epsilon_2(\ER) G_T(\ER, \Ed)$. The resulting rate within a detected energy interval $[ \Ed_1, \Ed_2]$ follows as
\beq
\label{R}
R_{[\Ed_1, \Ed_2]}(t) =
\frac{\rho}{\mDM} \sum_T \frac{C_T}{m_T} \int_0^\infty \ud \ER \, \int_{v \geqslant v_\text{min}(\ER)} \hspace{-18pt} \ud^3 v \, f(\bfv, t) \, v \, \frac{\ud \sigma_T}{\ud \ER}(\ER, \bfv)
\, \epsilon_2(\ER) \int_{\Ed_1}^{\Ed_2} \ud\Ed \, \epsilon_1(\Ed) G_T(\ER, \Ed) \ .
\eeq
The time dependence of the rate is generally well approximated by the first terms of a harmonic series,
\beq\label{Rt}
R_{[\Ed_1, \Ed_2]}(t) = R^0_{[\Ed_1, \Ed_2]} + R^1_{[\Ed_1, \Ed_2]} \cos\!\left[ \omega (t - t_0) \right] \ ,
\eeq
where $t_0$ is the time of the maximum of the signal and $\omega = 2 \pi/$yr. The coefficients $R^0_{[\Ed_1, \Ed_2]}$ and $R^1_{[\Ed_1, \Ed_2]}$ are, respectively, the unmodulated and modulated components of the rate in the energy interval $[\Ed_1, \Ed_2]$.

The differential cross section for the usual SI interaction is
\begin{align}
\label{dsigma_T}
\frac{\ud \sigma_T}{\ud \ER} = \sigma_T^{\rm SI}(\ER) \frac{m_T}{2 \mu_T^2 v^2} \ ,
&&
{\rm with}
&&
\sigma_T^{\rm SI}(\ER) = \sigma_p \frac{\mu_T^2}{\mu_p^2} [ Z_T + (A_T - Z_T) f_n / f_p ]^2 F_{{\rm SI}, T}^2(\ER) \ .
\end{align}
Here $Z_T$ and $A_T$ are respectively the atomic and mass number of the target nuclide $T$, $F_{{\rm SI}, T}(\ER)$ is the nuclear spin-independent form factor (which we take to be the Helm form factor \cite{Helm:1956zz} normalized to $F_{{\rm SI}, T}(0) = 1$), $f_n$ and $f_p$ are the effective WIMP couplings to neutron and proton, and $\mu_p$ is the WIMP-proton reduced mass. The WIMP-proton cross section $\sigma_p$ is the parameter customarily chosen to be constrained together with the WIMP mass $m$ for SI interactions, as it does not depend on the detector and thus bounds and allowed regions from different experiments can be compared on the same plot.

The isospin-conserving coupling  $f_n = f_p$ is  usually assumed by the experimental collaborations. The isospin-violating  coupling  $f_n / f_p = -0.7$ \cite{Kurylov:2003ra, Feng:2011vu}  produces the maximum  cancellation in the expression inside the square bracket in eq.~\eqref{dsigma_T} for  xenon, thus highly suppressing the interaction cross section. This suppression is phenomenologically interesting because it weakens considerably the bounds from xenon-based detectors such as XENON and LUX which provide some of the most restrictive bounds.

\medskip

In the SHM assumption, the DM velocity distribution in the galactic reference frame is a truncated Maxwell-Boltzmann \cite{Savage:2008er},
\beq
f_\text{G}(\bol{u}) = \frac{e^{- \bol{u}^2 / v_0^2}}{(v_0 \sqrt{\pi})^3 N_{\rm esc}} \Theta(\vesc - |\bol{u}|) \ ,
\eeq
with $N_{\rm esc}$ a constant so that $f_\text{G}$ is normalized to 1. The distribution can be boosted to Earth's reference frame with the Galilean transformation $f(\bol{v}, t) = f_\text{G}(\bol{u} = \bfv + \bfv_{\rm E}(t))$, where $\bfv$ is the velocity of the WIMP with respect to Earth and $\bfv_{\rm E}$ is the velocity of Earth with respect to the galaxy, whose average value is $\bfv_\odot$. We use $|\bfv_\odot|=232$ km/s \cite{Savage:2008er}, ${v}_0=220$ km/s, and ${v}_{\rm esc}=544$ km/s \cite{Smith:2006ym} (see \eg \cite{DelNobile:2013gba} for a discussion of the uncertainty on these parameters and its effect on the interpretation of the data). For the local DM density we use $\rho = 0.3$ GeV/$c^2$/cm$^3$.

\begin{figure}[t]
\centering
\includegraphics[width=0.49\textwidth]{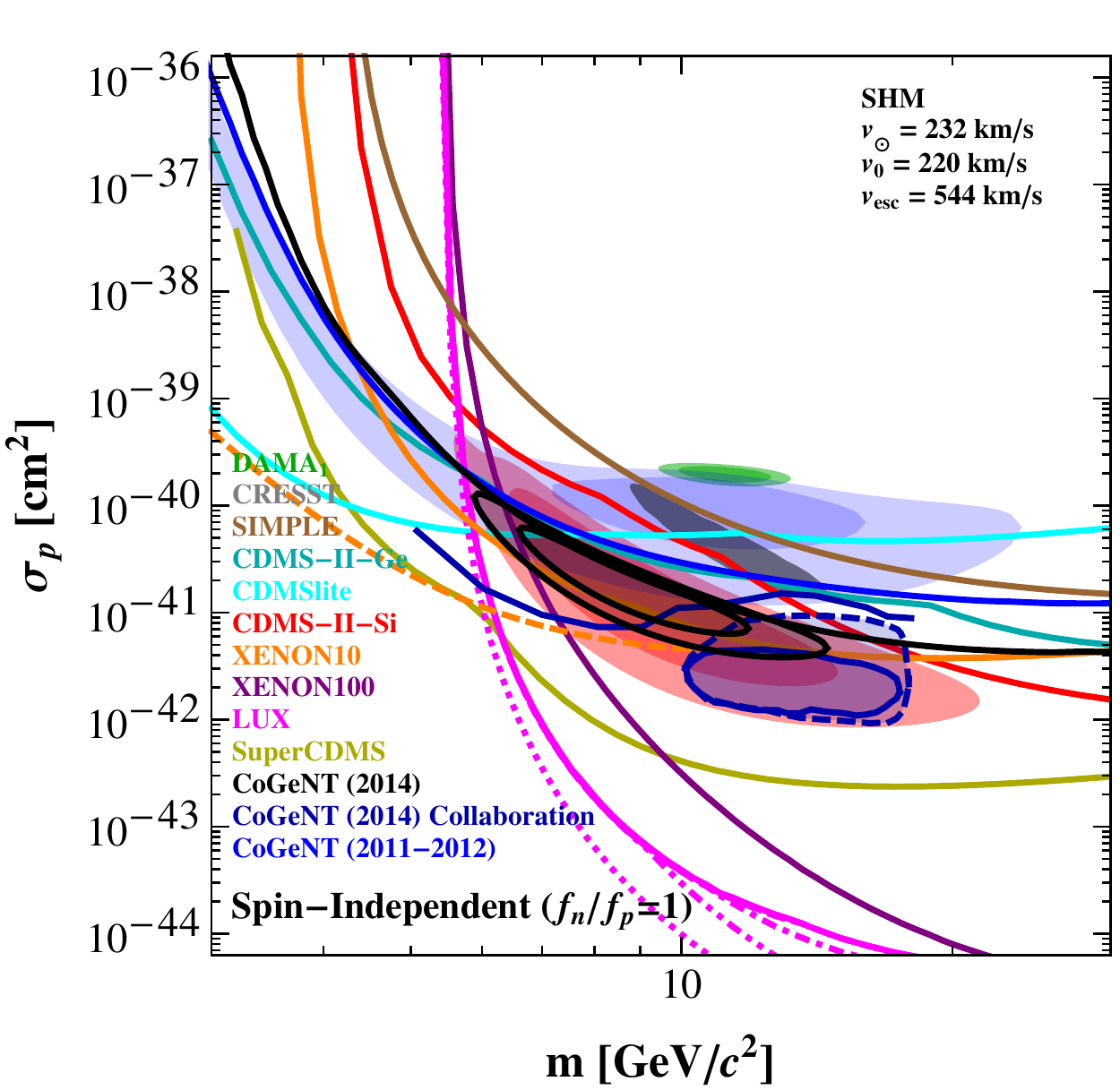}
\includegraphics[width=0.49\textwidth]{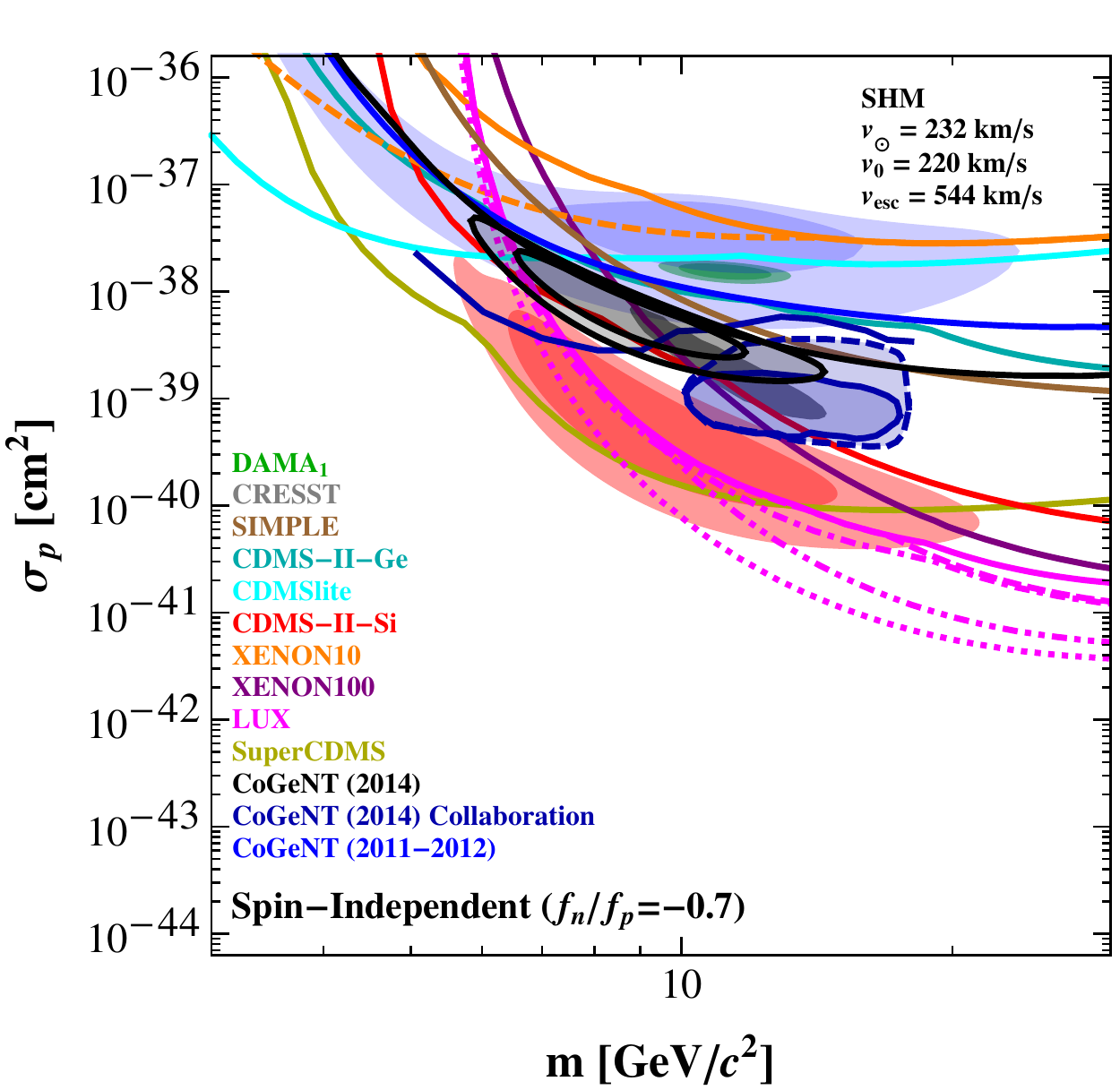}
\caption{$90\%$ CL bounds and $68\%$ and $90\%$ CL allowed regions in the SI DM-proton cross section vs WIMP mass plane, assuming the SHM. The CRESST-II low mass allowed region, from \cite{Angloher:2011uu}, is only shown at $2 \sigma$ CL. The blue line and filled light blue regions correspond to the upper limit on the unmodulated rate and the modulation amplitude region of the 2011-2012 CoGeNT data, respectively. The black line and contours show our CoGeNT 2014 bound and allowed region, while the dark blue limit line and the solid and dashed contours ($90\%$ regions with fixed and floating surface event background energy distributions, respectively) are those computed by the collaboration in \cite{Aalseth:2014jpa}. The left panel is for isospin-conserving $f_n = f_p$ couplings, the right panel is for isospin-violating $f_n / f_p = -0.7$ couplings. For XENON10 (orange bounds), the solid line is produced by conservatively setting the electron yield $\mathcal{Q}_{\rm y}$ to zero below 1.4 keVnr as in \cite{Aprile:2012nq}, while the dashed line ignores the $\mathcal{Q}_{\rm y}$ cut. For LUX (magenta bounds), the limits correspond (from the bottom) to 0, 1, 3, 5, and 24 observed events (see \cite{DelNobile:2013gba}). Only sodium is considered for DAMA, and the quenching factor is taken to be $Q_{\rm Na} = 0.3$.}
\label{m-sigma}
\end{figure}

Fig.~\ref{m-sigma} shows the SHM bounds and allowed regions in the $m$--$\sigma_p$ plane for both isospin-conserving (left panel) and isospin-violating interactions (right panel). We obtained the $90\%$ CL CDMS-II-Ge, CDMS-II-Si, CDMSlite, SuperCDMS, XENON10, XENON100 and LUX upper limits using the Maximum Gap Method \cite{Yellin:2002xd}. The SIMPLE bound is derived as the 90\% CL Poisson limit and the CoGeNT 2011-2012 unmodulated rate bound is the $90\%$ CL limit (in a raster scan). For isospin-conserving couplings all of the regions allowed by DAMA, CoGeNT, CRESST-II and CDMS-II-Si are entirely ruled out by the limits imposed by the null experiments, and for isospin-violating couplings only part of the CDMS-II-Si $90\%$ CL allowed region is compatible with all the bounds.

\section{Halo-independent analysis}
Using the differential cross section in eq.~\eqref{dsigma_T}, and changing integration variable from $\ER$ to $\vmin$ through eq.~\eqref{vmin}, we can rewrite the rate \eqref{R} as
\begin{align}
\label{R1}
R^{\rm SI}_{[\Ed_1, \Ed_2]}(t) & =  \int_0^\infty \ud \vmin \, \tilde{\eta}(\vmin, t) \, \eR^{\rm SI}_{[\Ed_1, \Ed_2]}(\vmin) \ ,
\end{align}
where the velocity integral $\tilde{\eta}$ is 
\beq
\label{eta0}
\tilde{\eta}(\vmin, t) \equiv \frac{\rho \sigma_p}{\mDM} \int_{v \geqslant \vmin} \ud^3 v \, \frac{f(\bfv, t)}{v} \equiv \int_{v \geqslant \vmin} \ud^3 v \, \frac{\tilde{f}(\bfv, t)}{v} \ ,
\eeq
and we defined the response function $\eR^{\rm SI}_{[\Ed_1, \Ed_2]}(\vmin)$ for WIMPs with SI interactions as
\beq
\label{Resp_SI}
\eR^{\rm SI}_{[\Ed_1, \Ed_2]}(\vmin) \equiv
2 \vmin \sum_T \frac{C_T}{m_T} \frac{\sigma_T^{\rm SI}(\ER(\vmin))}{\sigma_p}
\, \epsilon_2(\ER(\vmin)) \int_{\Ed_1}^{\Ed_2} \ud\Ed \, \epsilon_1(\Ed) G_T(\ER(\vmin), \Ed) \ .
\eeq
The velocity integral $\tilde{\eta}(\vmin,t)$ has an annual modulation due to Earth's rotation around the Sun, and can be separated into its unmodulated and modulated components as was done for the rate in eq.~\eqref{Rt},
\beq\label{etat}
\tilde{\eta}(\vmin, t) \simeq \tilde{\eta}^0(\vmin) + \tilde{\eta}^1(\vmin) \cos\!\left[ \omega (t - t_0) \right] .
\eeq
Once the WIMP mass and interactions are fixed, the functions $\tilde{\eta}^0(\vmin)$ and $\tilde{\eta}^1(\vmin)$ are detector-independent quantities that must be common to all experiments. Thus we can map the rate measurements and bounds of different experiments  into measurements of and bounds on $\tilde{\eta}^0(\vmin)$ and $\tilde{\eta}^1(\vmin)$ as functions of $\vmin$.

\medskip

For experiments with putative DM signals, in light of eq.~\eqref{R1} we may interpret the measured rates $\hat{R}^{\, i}_{[\Ed_1, \Ed_2]} \pm \Delta{R}^{\, i}_{[\Ed_1, \Ed_2]}$ in an energy interval $[\Ed_1, \Ed_2]$ as averages of the $\tilde{\eta}^i(\vmin)$ functions weighted by the response function $\eR^{\rm SI}_{[\Ed_1, \Ed_2]}(\vmin)$:
\beq
\label{avereta}
\overline{\tilde{\eta}^{\, i}_{[\Ed_1, \Ed_2]}} \equiv \frac{\hat{R}^{\, i}_{[\Ed_1, \Ed_2]}}
{\int \ud\vmin \, \eR^{\rm SI}_{[\Ed_1, \Ed_2]}(\vmin)} \ ,
\eeq
with $i = 0, 1$ for the unmodulated and modulated component, respectively. Each such average corresponds to a point with error bars in the $(\vmin, \tilde{\eta})$ plane. The vertical bars  are given by $\Delta\overline{\tilde{\eta}^{\, i}_{[\Ed_1, \Ed_2]}}$ computed by replacing $\hat{R}^{\, i}_{[\Ed_1, \Ed_2]}$ with $\Delta{R}^{\, i}_{[\Ed_1, \Ed_2]}$  in eq.~\eqref{avereta}.  The  $\Delta{R}^{\, i}$ used here correspond to  the $68\%$ confidence interval. The horizontal bar shows the $\vmin$ interval where the response function $\eR^{\rm SI}_{[\Ed_1, \Ed_2]}(\vmin)$ for the given experiment is sufficiently different from zero. Following \cite{Frandsen:2011gi, Gondolo:2012rs, DelNobile:2013cta, DelNobile:2013gba} the horizontal bar may be chosen to extend over the interval $[{\vmin}_{,1}, {\vmin}_{,2}] = [v_{\rm min}(\Ed_1 - \sigma_{\ER}(\Ed_1)), v_{\rm min}(\Ed_2 + \sigma_{\ER}(\Ed_2))]$, where $\sigma_{\ER}(\Ed)$ is the energy resolution and the function $v_{\rm min}(\Ed)$ is obtained from $v_{\rm min}(\ER)$ in eq.~\eqref{vmin} by using the recoil energy  $\ER$ that produces the mean $\langle \Ed \rangle$ which is equal to the measured energy $\Ed$. When isotopes of the same element are present, like for Xe or Ge, the $v_{\rm min}$ intervals of the different isotopes almost completely overlap, and we take $v_{\rm min,1}$ and $v_{\rm min,2}$ to be the $C_T$-weighted averages over the isotopes of the element. When there are nuclides belonging to very different elements, like Ca and O in CRESST-II, a more complicated procedure should be followed (see \cite{Frandsen:2011gi, Gondolo:2012rs} for details).

To determine the upper bounds on the unmodulated part of $\tilde{\eta}$, a procedure first outlined in \cite{Fox:2010bz, Frandsen:2011gi} may be used.  This procedure exploits the fact that, by definition, $\tilde{\eta}^0$ is a non-increasing function of $\vmin$. For this reason, the smallest possible $\tilde{\eta}^0(\vmin)$ function passing by a fixed point $(v_0, \tilde{\eta}_0)$ in the $(\vmin, \tilde{\eta})$ plane is the downward step-function $\tilde{\eta}_0 \, \theta(v_0 - \vmin)$.  In other words, among the functions passing by the point $(v_0, \tilde{\eta}_0)$, the downward step is the function yielding the minimum predicted number of events. Imposing this functional form in eq.~\eqref{R1} we obtain
\beq
R_{[\Ed_1, \Ed_2]} = \tilde{\eta}_0 \int_0^{v_0} \ud \vmin \, \eR^{\rm SI}_{[\Ed_1, \Ed_2]}(\vmin) \ .
\eeq
The upper bound $R^{\rm lim}_{[\Ed_1, \Ed_2]}$ on the unmodulated rate in an interval $[\Ed_1, \Ed_2]$ is translated into an upper bound $\tilde{\eta}^{\rm lim}(\vmin)$ on $\tilde{\eta}^0$ at $v_0$ by
\beq
\tilde{\eta}^{\rm lim}(v_0) = \frac{R^{\rm lim}_{[\Ed_1, \Ed_2]}}{\int_0^{v_0} \ud \vmin \, \eR^{\rm SI}_{[\Ed_1, \Ed_2]}(\vmin)} \ .
\eeq
The upper bound so obtained is conservative in the sense that any $\tilde{\eta}^0$ function extending even partially above $\tilde{\eta}^{\rm lim}$ is excluded, but not every $\tilde{\eta}^0$ function lying everywhere below $\tilde{\eta}^{\rm lim}$ is allowed \cite{Frandsen:2011gi}.

\medskip

The procedure just described does not assume any particular property of the DM halo. By making some assumptions, more stringent limits on the modulated part $\tilde{\eta}^1$ can be derived  from the limits on the unmodulated part of the rate (see \cite{Frandsen:2011gi, HerreroGarcia:2011aa, HerreroGarcia:2012fu, Bozorgnia:2013hsa}), but we choose to proceed without making any assumption on the DM halo.

\medskip

\begin{figure}[!htbp]
\centering
\includegraphics[width=0.46\textwidth]{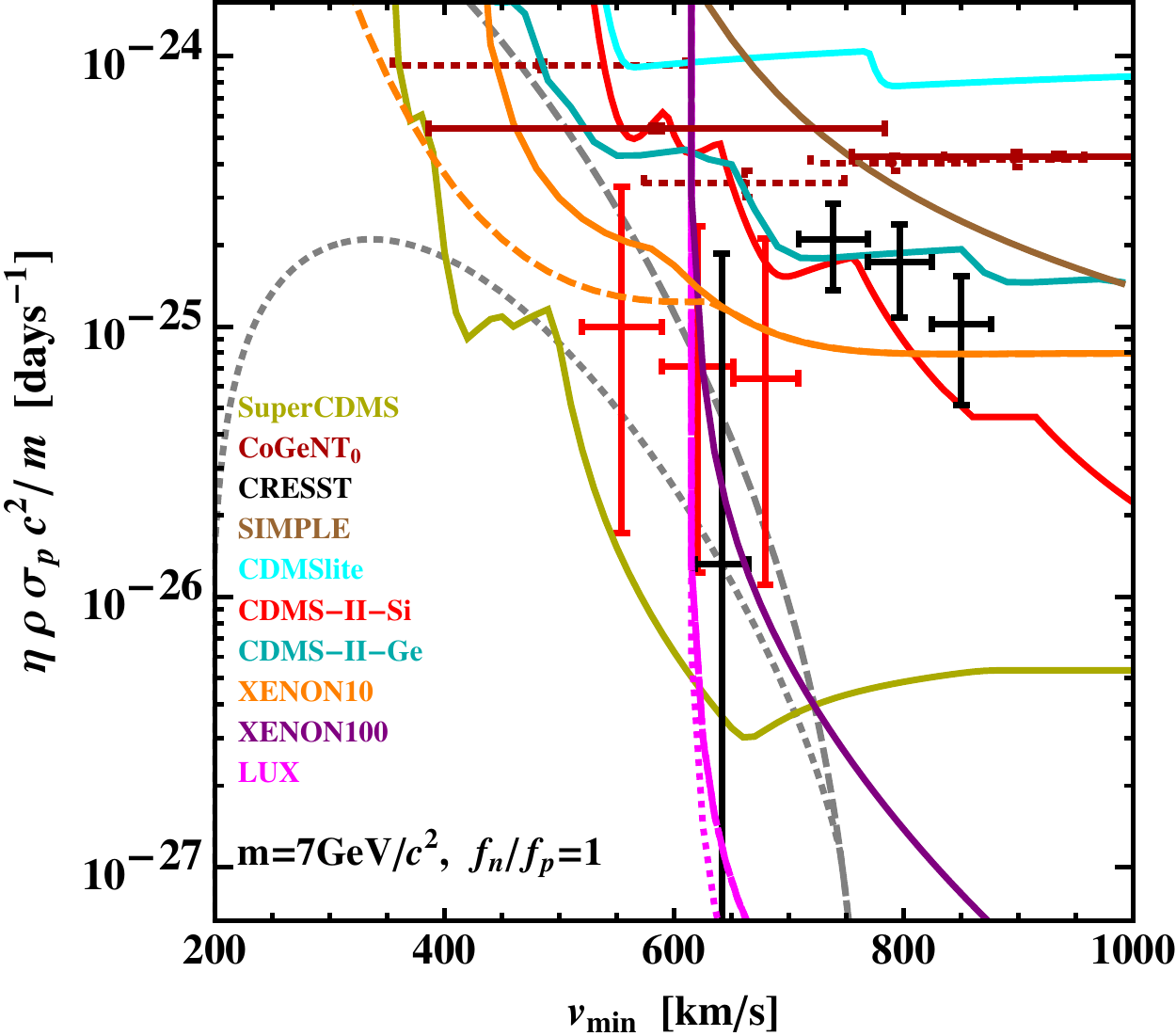}
\includegraphics[width=0.46\textwidth]{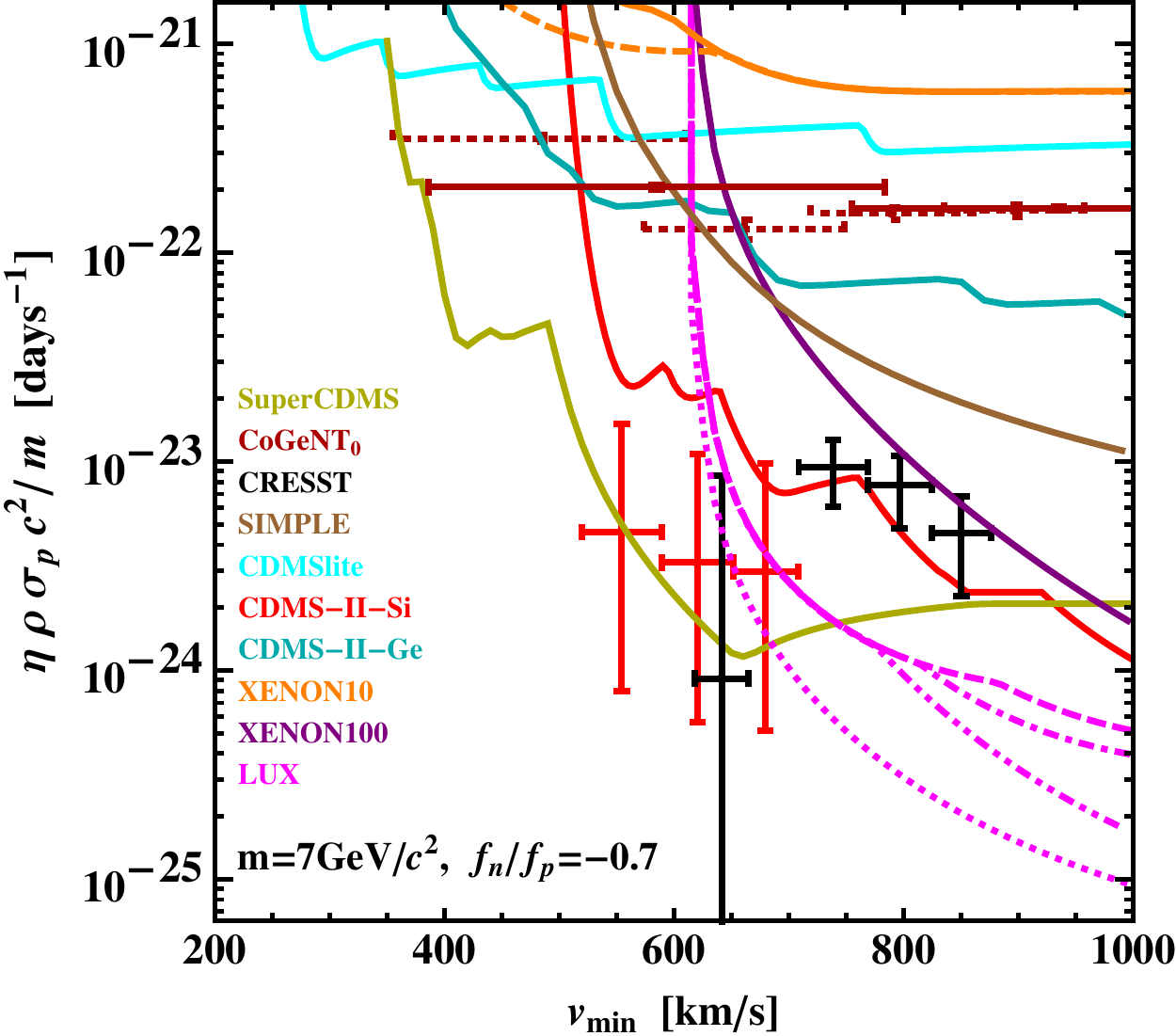}
\\
\includegraphics[width=0.46\textwidth]{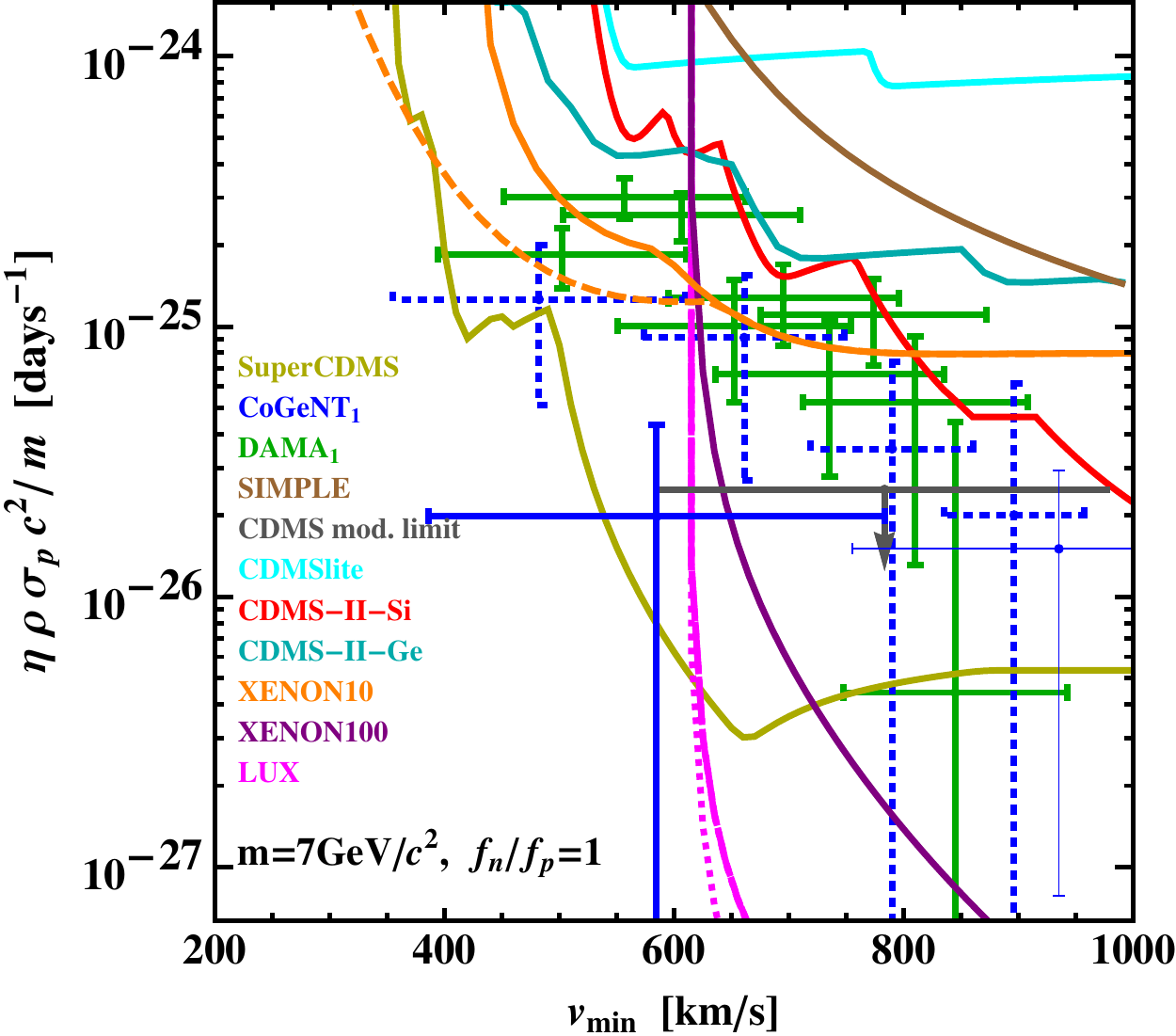}
\includegraphics[width=0.46\textwidth]{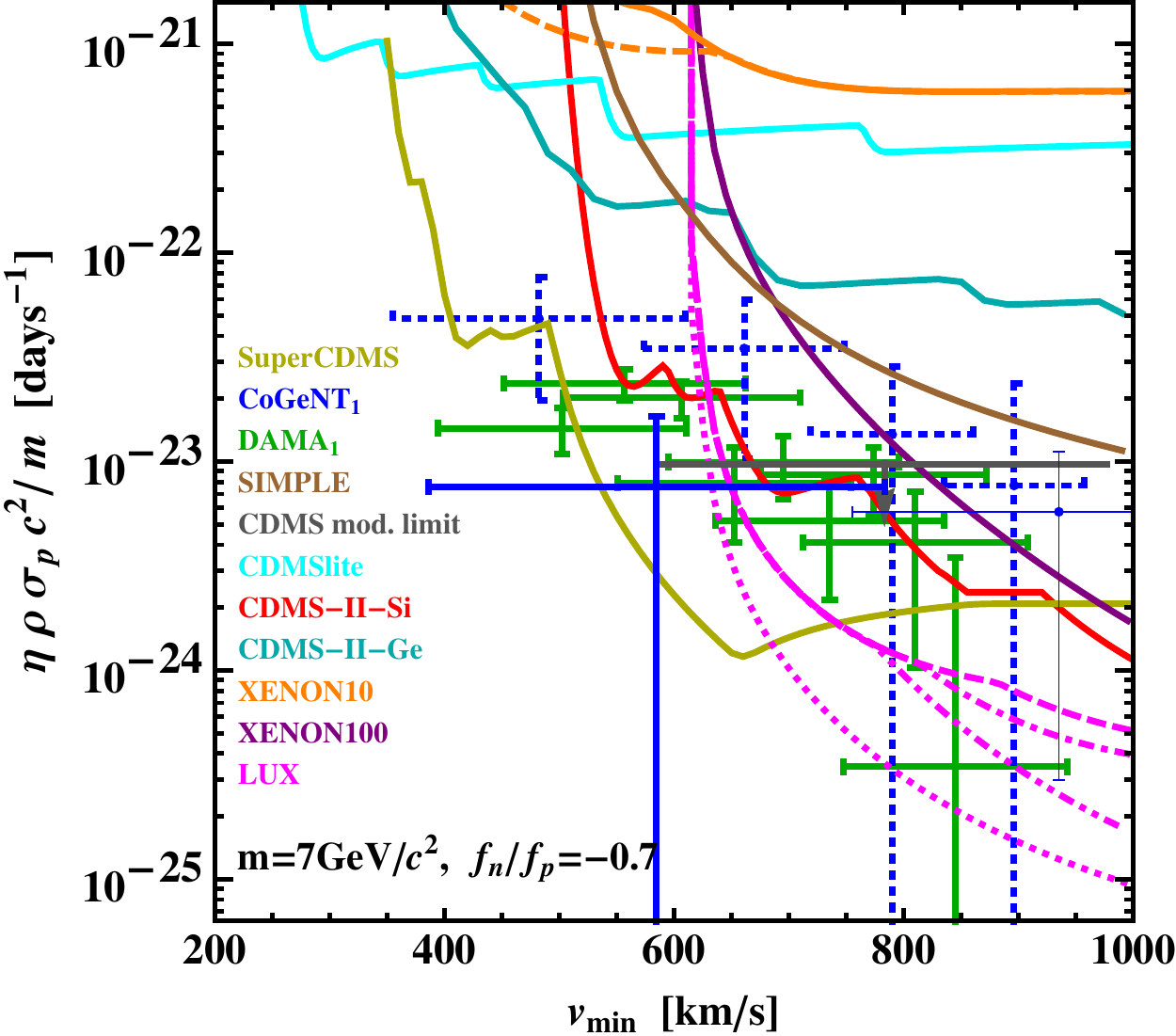}
\\
\includegraphics[width=0.46\textwidth]{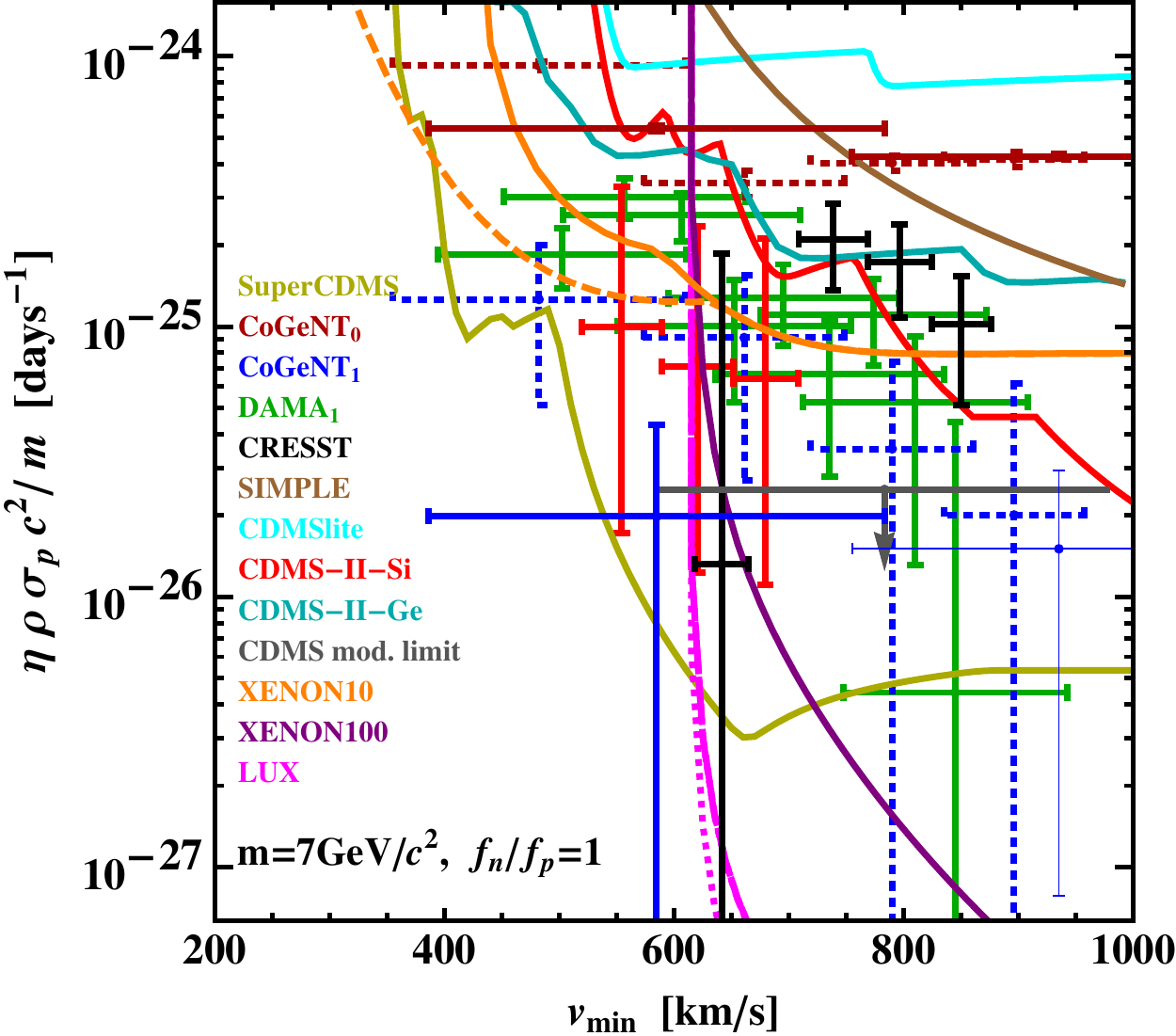}
\includegraphics[width=0.46\textwidth]{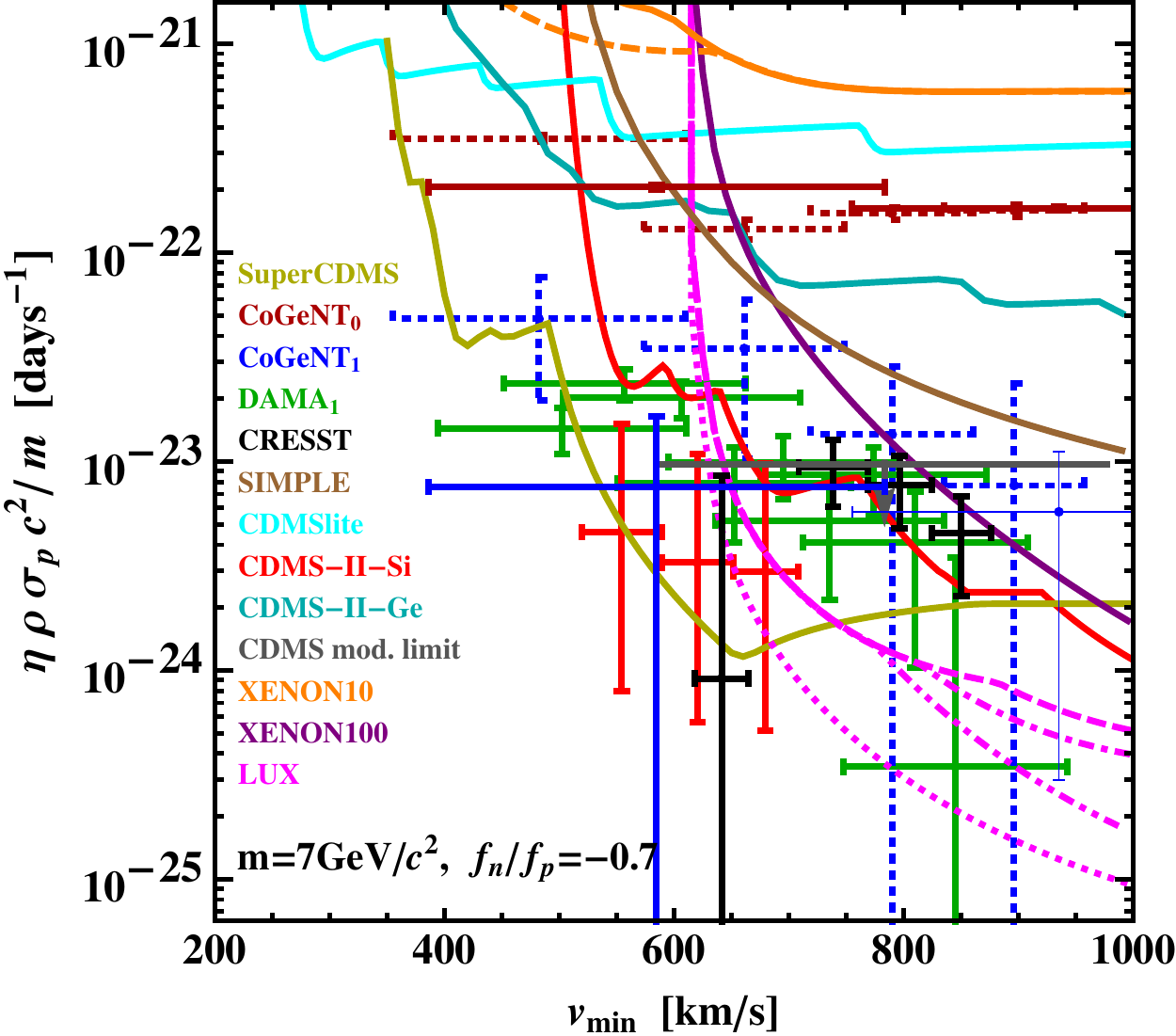}
\caption{Measurements of and bounds on $\tilde\eta c^2$ for $m = 7$ GeV/$c^2$. The left and right columns are for isospin-conserving  and isospin-violating interactions, respectively. The top, middle and bottom rows show measurements and bounds for the unmodulated component $\tilde\eta^0 c^2$, for the modulated component $\tilde\eta^1 c^2$, and for both together, respectively. The middle row also shows the upper bounds on $\tilde\eta^0 c^2$ from the plots on the top row. The dashed gray lines in the top left panel show the SHM $\tilde{\eta}^0 c^2$ (upper line) and $\tilde{\eta}^1 c^2$ (lower line) for $\sigma_p = 1 \times 10^{-40}$ cm$^2$, which provides a good fit to the CDMS-II-Si data.
}
\label{eta7}
\end{figure}

\begin{figure}[!htbp]
\centering
\includegraphics[width=0.46\textwidth]{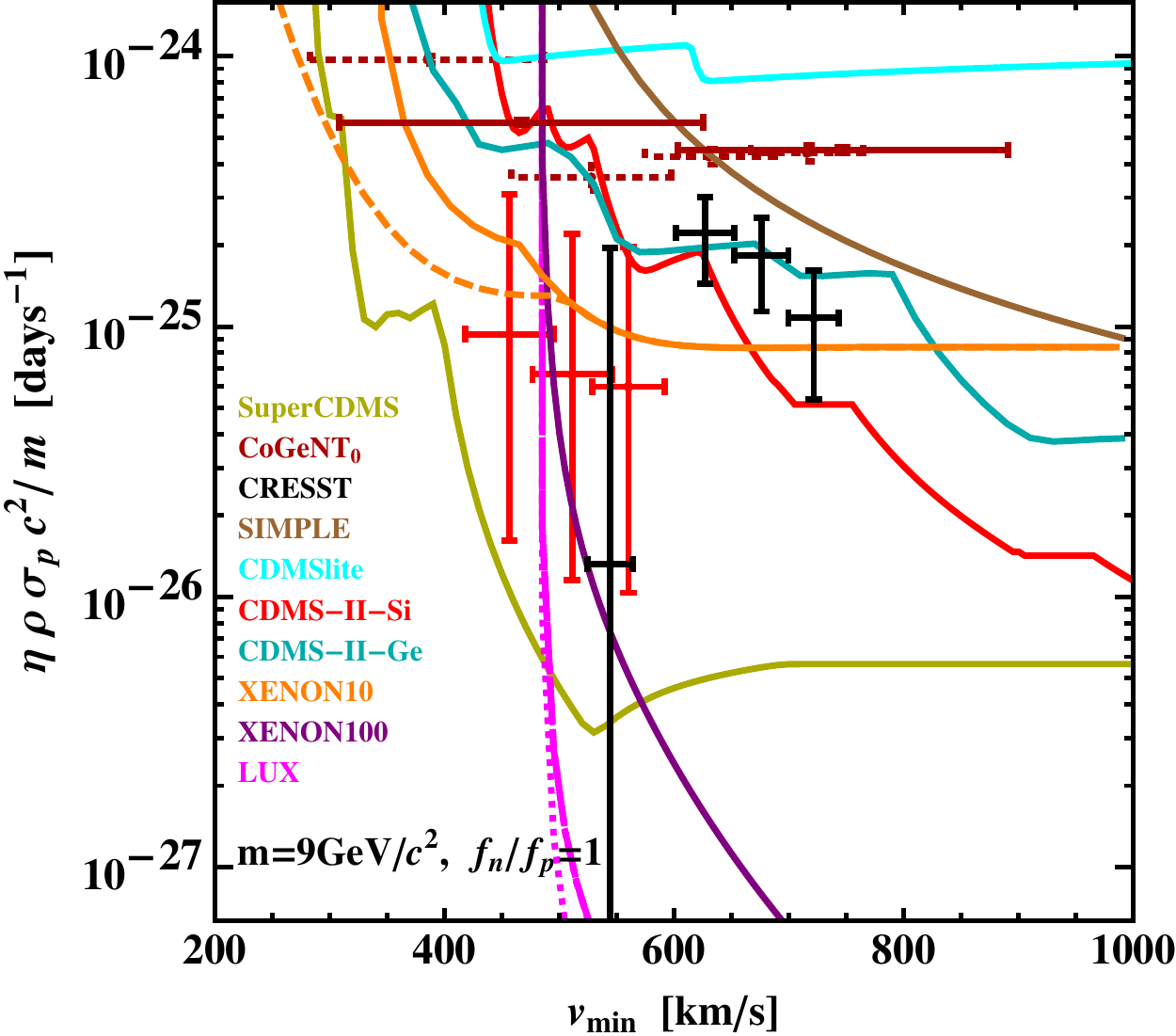}
\includegraphics[width=0.46\textwidth]{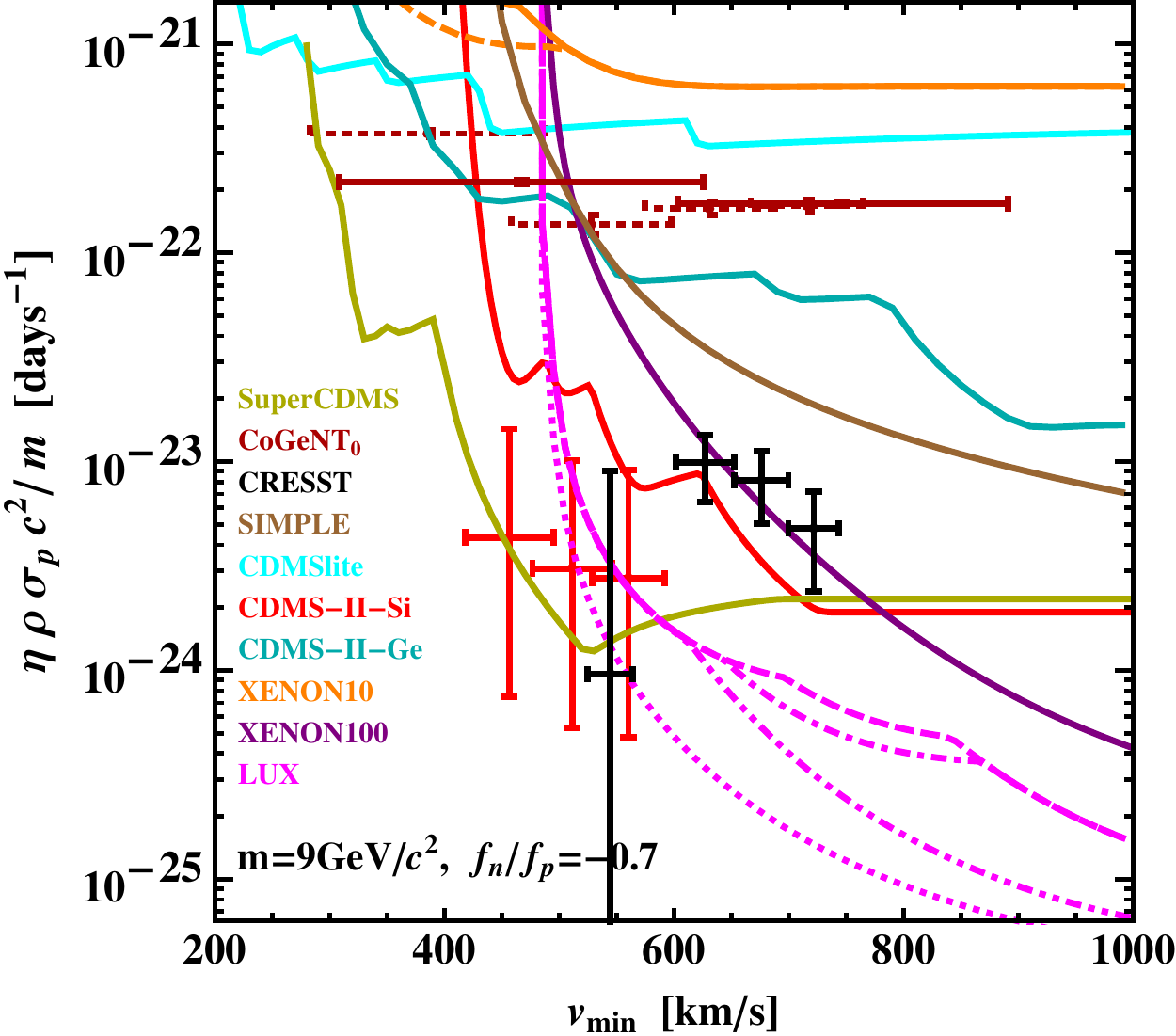}
\\
\includegraphics[width=0.46\textwidth]{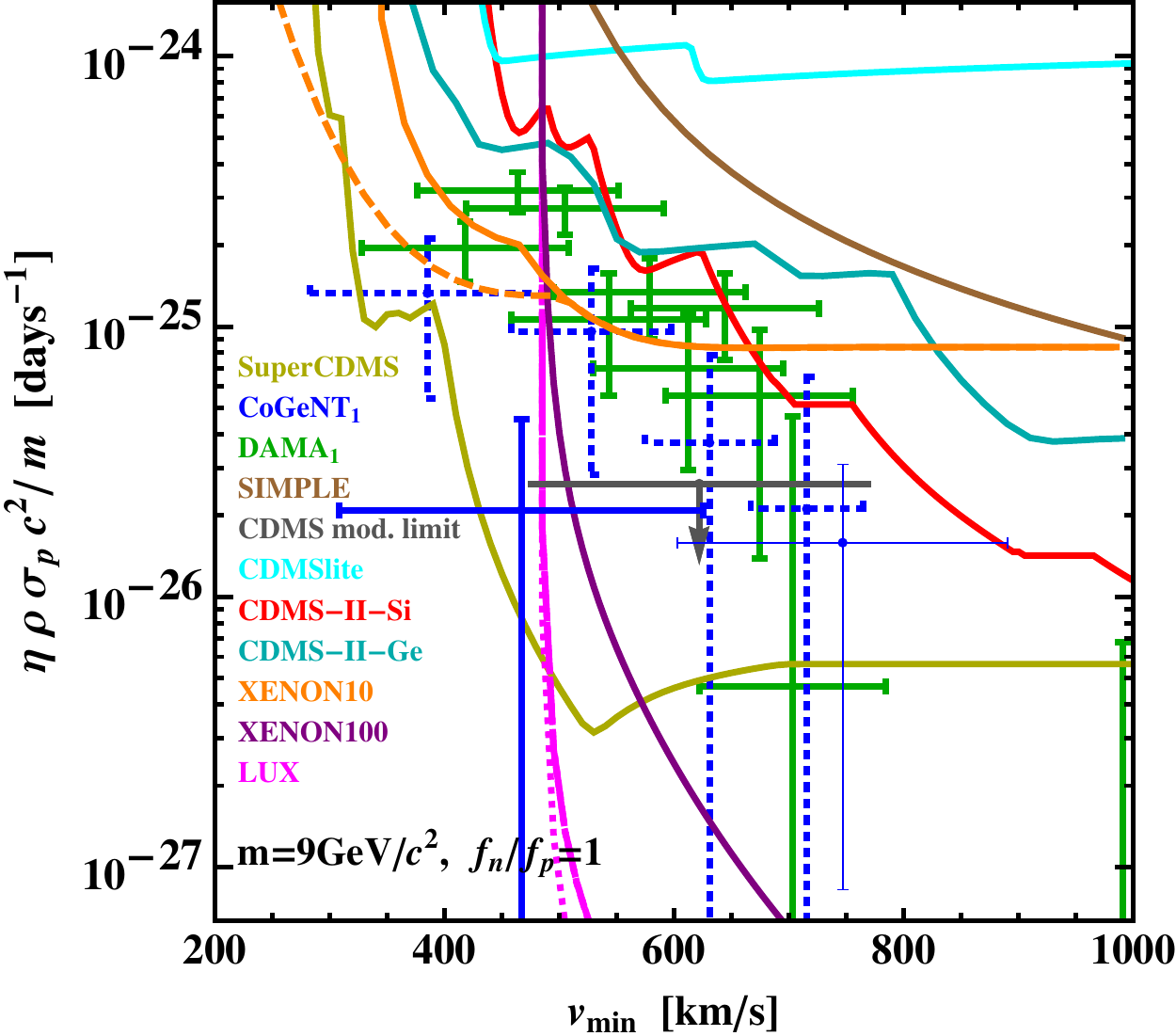}
\includegraphics[width=0.46\textwidth]{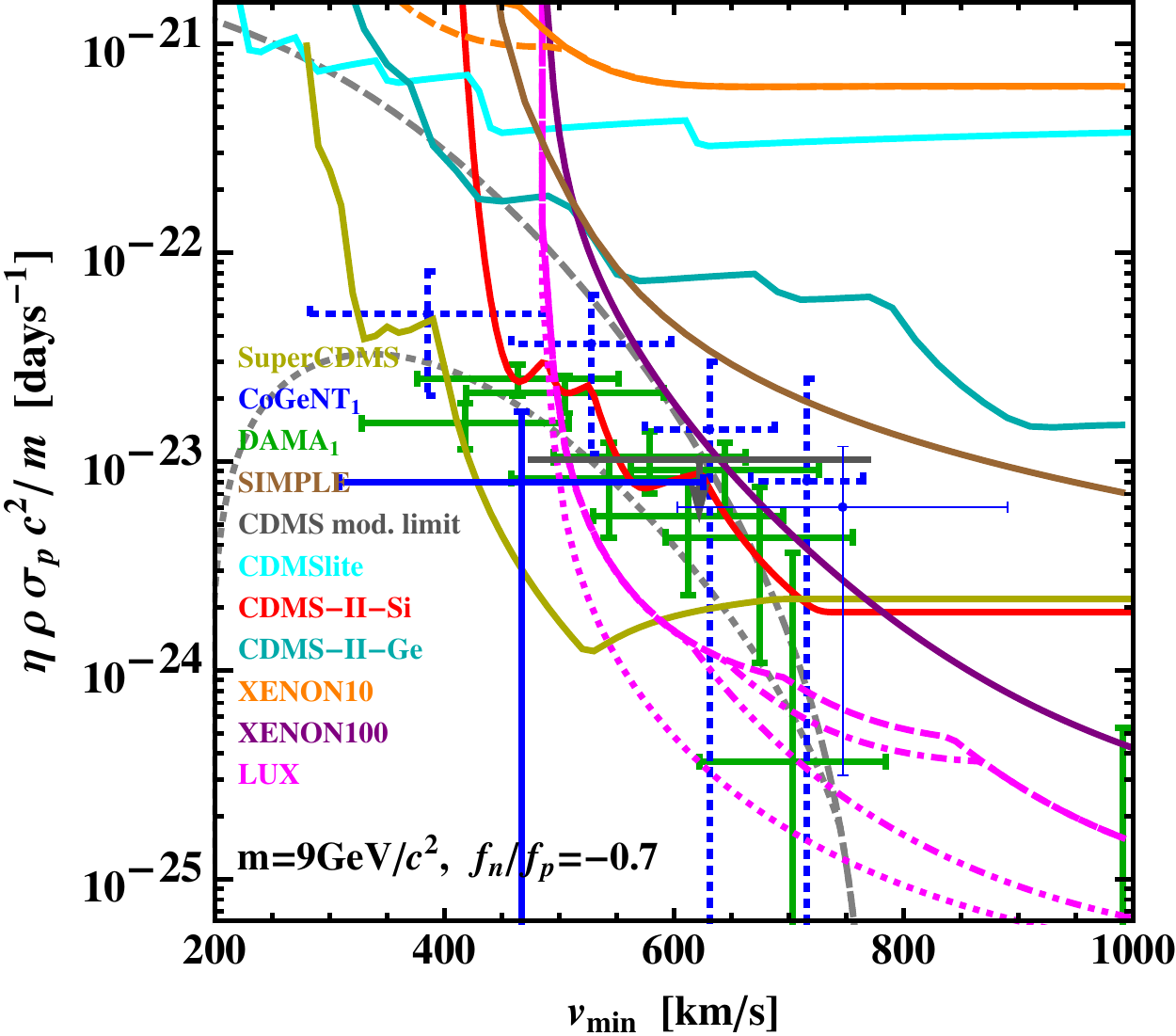}
\\
\includegraphics[width=0.46\textwidth]{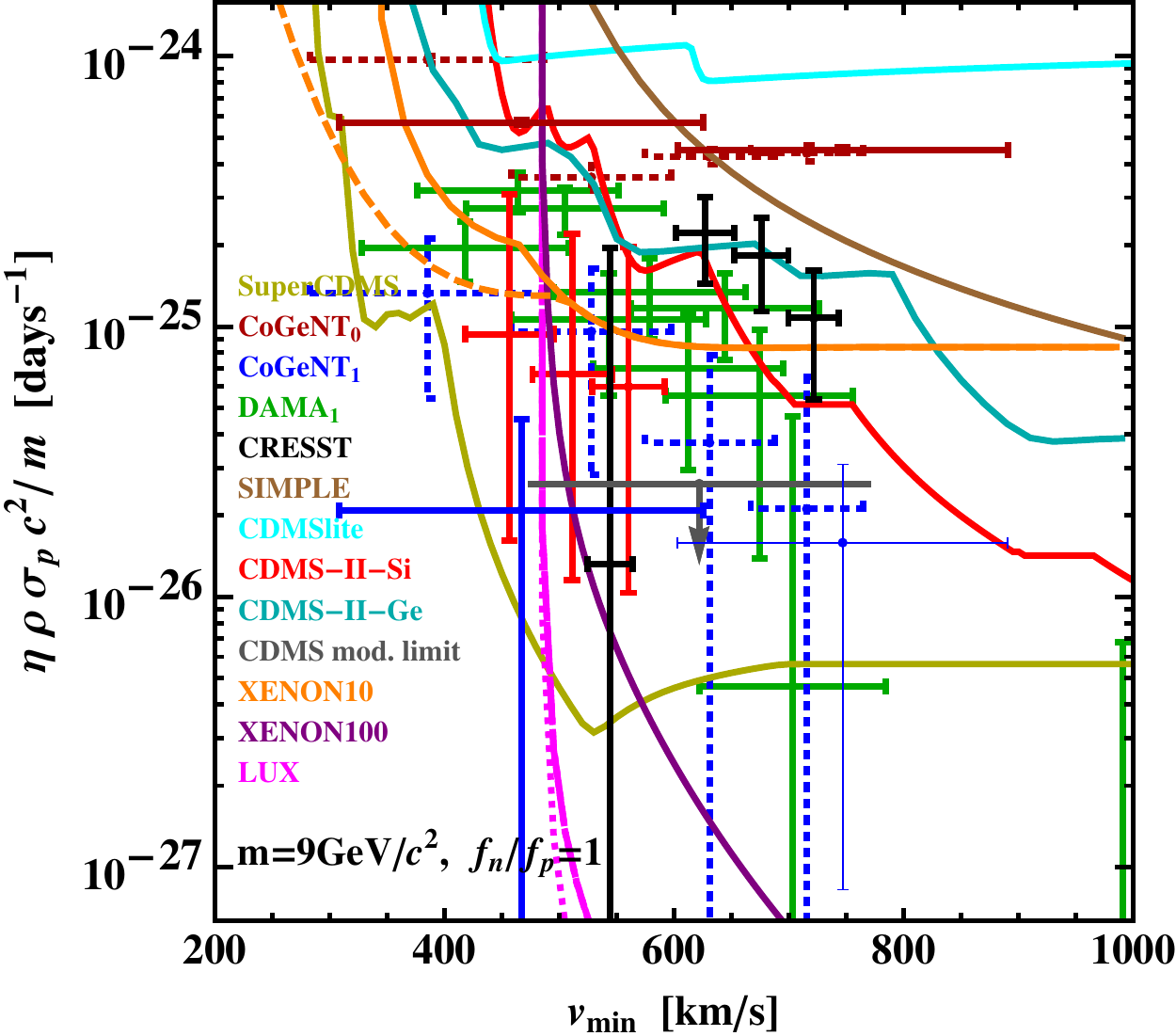}
\includegraphics[width=0.46\textwidth]{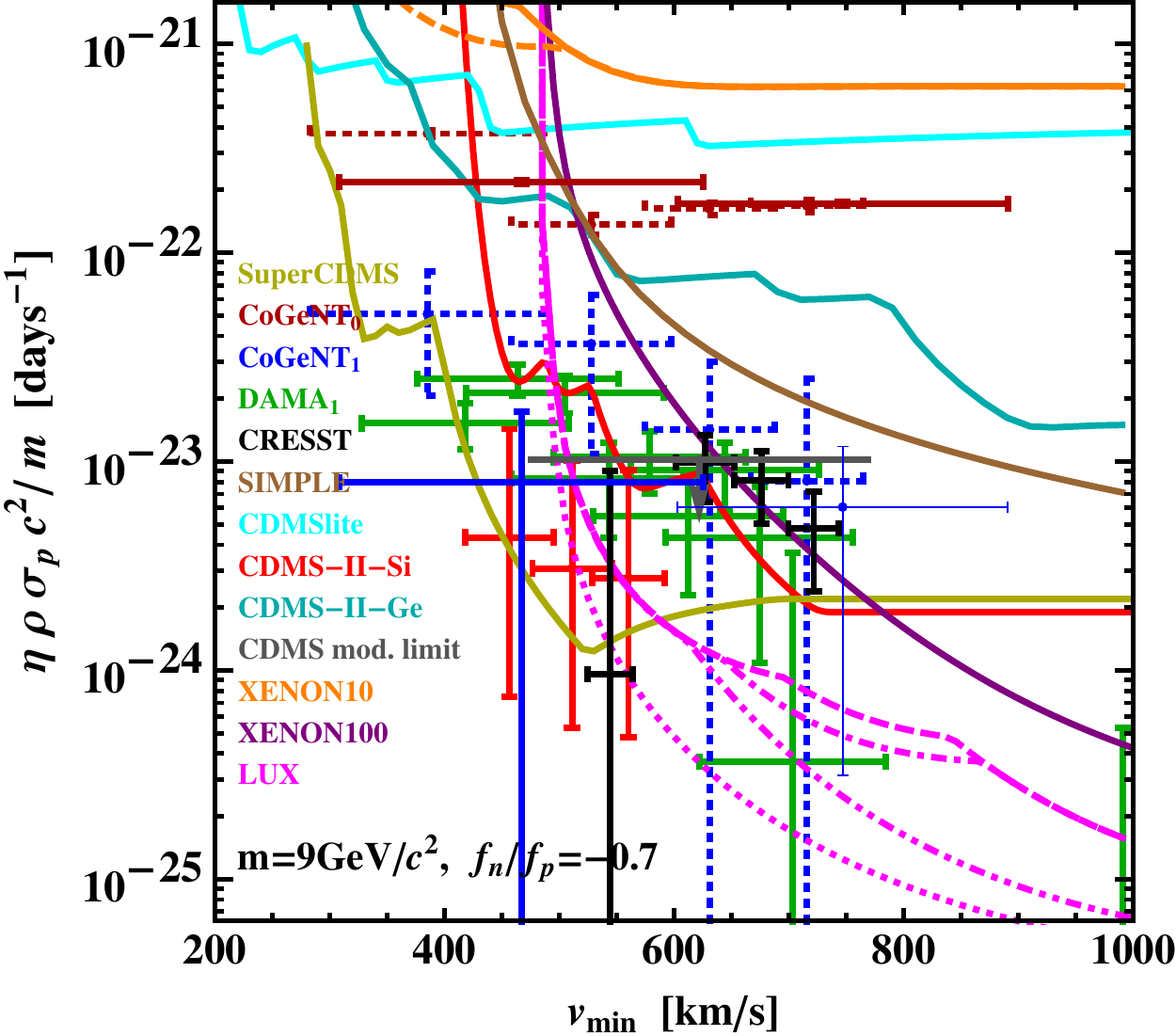}
\caption{Same as fig.~\ref{eta7}, but for a DM mass $m = 9$ GeV/$c^2$. The dashed gray lines in the middle right panel show the expected $\tilde{\eta}^0 c^2$ (upper line) and $\tilde{\eta}^1 c^2$ (lower line) for a WIMP-proton cross section $\sigma_p = 2 \times 10^{-38}$ cm$^2$ in the SHM, which gives a good fit to the DAMA modulation data.}
\label{eta9}
\end{figure}

Figs.~\ref{eta7} and \ref{eta9} collect the results of the halo-independent analysis for a WIMP mass $m = 7$ GeV/$c^2$ and $m = 9$ GeV/$c^2$, respectively. In each figure, to avoid cluttering, we separate the material as follows: the top row shows measurements of and bounds on $\tilde{\eta}^0 c^2$; the middle row shows measurements of and the CDMS-II-Ge limit on $\tilde{\eta}^1 c^2$, plus the same bounds on $\tilde{\eta}^0 c^2$ of the top row; and the bottom row shows all measurements and bounds together. The left and right columns correspond to isospin-conserving ($f_n = f_p$) and isospin-violating ($f_n / f_p = -0.7$) interactions, respectively. The crosses show the DAMA modulation signal (green crosses), and the CRESST-II and CDMS-II-Si unmodulated signals (black and red crosses, respectively). For CoGeNT, the blue and dark red crosses (with very small vertical error bars) show the modulated signal and unmodulated signal (plus an unknown flat background), respectively; the solid crosses are for the 2014 data set (we indicate with thin lines the modulus of the negative part of each cross), while the dashed crosses are for the 2011-2012 data set alone. The CDMS-II-Ge modulation bound is shown as a dark grey horizontal line with downward arrow. As in the SHM analysis presented in fig.~\ref{m-sigma}, only sodium is considered for DAMA (with quenching factor $Q_{\rm Na} = 0.3$), as for the DM masses considered here the WIMP scattering off iodine is supposed to be below threshold. For XENON10, limits produced by setting or not setting the electron yield $\mathcal{Q}_{\rm y}$ to zero below 1.4 keVnr (as in \cite{Angle:2011th}) are obtained (solid and dashed orange line, respectively). For LUX, upper bounds considering 0, 1, 3 and 5 observed events are computed \cite{DelNobile:2013gba}, corresponding (from bottom to top) to the magenta lines with different dashing styles.

The overlapping of the green and dashed blue crosses in figs.~\ref{eta7} and \ref{eta9} seems to indicate that the DAMA and CoGeNT 2011-2012 modulation rates are compatible with each other; however, the solid blue CoGeNT 2014 crosses lie in general below the DAMA data and are in fact compatible with zero modulation at about $1 \sigma$ \cite{DelNobile:2013gba}. Moreover, the three CDMS-II-Si points overlap or are below the CoGeNT and DAMA measurements of the modulated part of $\tilde{\eta}$. Thus, interpreted as a measurement of the unmodulated rate, the three CDMS-II-Si data points seem largely incompatible with the modulation of the signal observed by CoGeNT and DAMA, since a modulated signal is expected to be much smaller than the respective unmodulated component. For isospin-conserving interactions (left column of figs.~\ref{eta7} and \ref{eta9}), the experiments with a positive signal seem largely incompatible with the limits set by the other experiments, most notably by SuperCDMS and LUX. The compatibility of the DAMA, CoGeNT and CRESST-II data with the exclusion bounds improves slightly for isospin-violating couplings with $f_n / f_p = -0.7$, for which the XENON and LUX limits are weakened (right column of figs.~\ref{eta7} and \ref{eta9}). However, only the CDMS crosses fall partially below the exclusion lines.

\section{Conclusions}

We have reviewed the halo-independent method to compare data from direct DM detection experiments, following closely the treatment in \cite{DelNobile:2013cta, DelNobile:2013cva, DelNobile:2013gba, DelNobile:2014eta}. We applied the halo-independent analysis to SI interactions with both isospin-conserving and isospin-violating couplings, updating on previous results with the addition of the SuperCDMS limit. For both choices of the coupling the situation seems to be of disagreement between most of the experiments with positive signals (DAMA, CoGeNT, CRESST-II) and those with negative results (most notably SuperCDMS and LUX). The three CDMS-II-Si events seem however compatible with all the limits for DM with isospin-violating couplings. DAMA and CoGeNT 2011-2012 modulation rates seem to agree with each other, but they appear to be incompatible with the CDMS-II-Si events when these are interpreted as measurements of the unmodulated rate. The CoGeNT 2014 modulation rates are instead compatible with zero at the $1 \sigma$ level.

The halo-independent analysis is a promising framework to compare different direct detection experiments without making assumptions on the DM halo. This feature is highly desirable given the crucial role played by the DM velocity distribution in the galaxy in determining the total scattering rate at direct detection experiments. This analysis allows to directly compare the recoil spectra measured by different experiments in $\vmin$ space, together with bounds from null experiments. These spectra indicate the integrated DM velocity distribution $\tilde{\eta}$ favored by the experiments, as a function of $\vmin$ (see eq.~\eqref{eta0}).

At present this framework, which can be used for any WIMP-nucleus interaction \cite{DelNobile:2013cva, DelNobile:2014eta}, presents some drawbacks which could be addressed and improved in future work (\eg see the recent attempts in \cite{Feldstein:2014gza, Fox:2014kua}). For instance, the relation between the $\tilde{\eta}$ function that one wants to fit and the observed rates is an integral equation, eq.~\eqref{R1}. So far we only computed the weighted average of $\tilde{\eta}(\vmin, t)$ in a $\vmin$ interval, eq.~\eqref{avereta}, and in general this is a poor representation of the $\tilde{\eta}$ function within this interval. Secondly, the degree of agreement or disagreement between two data sets can not be statistically quantified in the current halo-independent analysis. Finally, although by making some (mild) assumptions, more stringent limits on the modulated part of the rate $\tilde{\eta}^1$ can be derived from the limits on the unmodulated part $\tilde{\eta}^0$ \cite{Frandsen:2011gi, HerreroGarcia:2011aa, HerreroGarcia:2012fu, Bozorgnia:2013hsa}, with no additional assumptions we can only use the most general inequality $\tilde{\eta}^0 > \tilde{\eta}^1$.

\section*{Acknowledgments}
%The author acknowledges partial support by Department of Energy under Award Number DE-SC0009937.
P.G.~was supported in part by NSF grant PHY-1068111. E.D.N., G.G.~and J.-H.H.~were supported in part by the Department of Energy under Award Number DE-SC0009937. J.-H.H.~was also partially supported by the Spanish Consolider-Ingenio MultiDark Project (CSD2009-00064).



%% References


\bibliographystyle{elsarticle-num}
\bibliography{bibliography}



%\begin{thebibliography}{00}
%
%%% \bibitem must have the following form:
%%%   \bibitem{key}...
%%%
%
%% \bibitem{}
%
%\end{thebibliography}

\end{document}